\documentclass[prl,twocolumn]{revtex4}
\usepackage{graphicx,epsf}
\usepackage{color}
\usepackage{bm}
\usepackage{dcolumn}
\usepackage{amsfonts}
\usepackage{mathrsfs}

\usepackage{subfig}

\begin{document}

\title{On scattering and damping of Toroidal Alfv\'en eigenmode by drift wave turbulence}

\author{ Liu Chen$^{1,2,3}$, Zhiyong Qiu$^{1,3}$, and Fulvio Zonca$^{3, 1}$}

\affiliation{$^1$Institute for    Fusion Theory and Simulation, School of Physics, Zhejiang University, Hangzhou, P.R.C\\
$^2$Department of   Physics and Astronomy,  University of California, Irvine CA 92697-4575, U.S.A.\\
$^3$ Center for Nonlinear Plasma Science and   C.R. ENEA Frascati, C.P. 65, 00044 Frascati, Italy}

\begin{abstract}
We demonstrate analytically that, in toroidal plasmas, scattering by drift wave turbulence could lead to appreciable damping of toroidal Alfv\'en eigenmodes via generation of short-wavelength electron Landau damped  kinetic Alfv\'en waves. A corresponding analytic expression of the damping rate is derived, and found to be, typically, comparable to the linear drive by energetic particles. The implications of this novel mechanism on the transport and heating processes in burning plasmas are also discussed.
\end{abstract}

%\pacs{52.30.Gz, 52.35.Bj,52.35.Fp, 52.35.Mw}

\maketitle

Microscopic drift wave  (DW) turbulences are intrinsic to magnetically confined plasmas such as tokamaks \cite{WHortonRMP1999}. These electrostatic waves have, typically, frequencies in the range of electron/ion diamagnetic drift frequencies and perpendicular wavelengths comparable to ion Larmor radii. On the other hand,  shear Alfv\'en waves (SAWs) or, more precisely, Alfv\'en eigenmodes (AEs)  excited by energetic particles (EPs) \cite{CZChengAP1985,AFasoliNF2007,LChenRMP2016}, have been observed in current experiments \cite{KWongPRL1991,WHeidbrinkPRL1993,SSharapovNF2013}, and predicted for future burning plasma experiments such as ITER \cite{KTomabechiNF1991}. Since these EP-driven AEs have direct bearing on the confinement of EPs (including fusion $\alpha$ particles) and, consequently, the plasma performance, it is crucial that the      stability properties be accurately assessed.
In this Letter,  we demonstrate analytically, for the first time, that ambient DWs could lead to appreciable  damping of toroidal Alfv\'en eigenmode (TAE) \cite{CZChengAP1985} by cross-scale couplings via direct  nonlinear wave-wave interactions.    Physically, this damping occurs because DWs scatter the TAE into short-wavelength kinetic Alfv\'en waves (KAWs) \cite{AHasegawaPoF1976} which are then Landau damped by, mainly, electrons.  An explicit  analytic expression of the DWs-induced  damping rate is derived and, for typical tokamak parameters, is found to be comparable to the EP-driven TAE growth rate. We also note that the novel nonlinear wave-wave interactions  carry significant implications to confinement and heating processes in burning plasmas.

Let's consider a tokamak plasma with circular magnetic surfaces and a large aspect ratio, i.e., $\epsilon\equiv r/R<1$. Here, $r$ and $R$ are, respectively, the minor and major radii of the torus. We, furthermore,  take $\beta\equiv P/B^2\sim O(\epsilon^2)\ll1$, the equilibrium distributions be Maxwellian, and the density be nonuniform. However, in order, as a paradigm model, to simplify the analysis, we have, for now, neglected the effects of finite temperature gradients and trapped particles. Thus, the present work is focused on the roles played by electron drift waves in scattering and damping of TAE. With $\beta\ll1$, magnetic compression can be neglected, and we may adopt $\delta\phi$ and $\delta A_{\parallel}$ as the field variables. Here, $\delta\phi$  is the scalar potential and $\delta A_{\parallel}$ is the parallel component of the vector potential; i.e., $\delta \mathbf{A} \simeq \delta A_{\parallel}\mathbf{b}$ and $\mathbf{b}\equiv \mathbf{B}_0/B_0$. The perturbed distribution function, $\delta f_j$ with $j=e,i$, then obeys the following nonlinear gyrokinetic equations \cite{EFriemanPoF1982}
\begin{eqnarray}
\delta f_j=-(e/T)_j\delta\phi F_{Mj} +\exp(-\bm{\rho}_j\cdot\nabla)\delta g_j,
\end{eqnarray}
and
\begin{eqnarray}
&&\left(\partial_t +v_{\parallel}\mathbf{b}\cdot\nabla +\mathbf{v}_d\cdot\nabla +\langle\delta \mathbf{u}_{gj}\rangle_{\alpha}\cdot\nabla \right)\delta g_j\nonumber\\
&=&(e/T)_jF_{Mj}(\partial_t+i\omega_{*j})\langle \exp(\bm{\rho}_j\cdot\nabla) \delta L \rangle_{\alpha}.\label{eq:nl_gk}
\end{eqnarray}
Here, $F_{Mj}$ is the Maxwellian distribution, $\Omega_j=(eB_0/mc)_j$, $\mathbf{v}_d=\mathbf{b}\times [(v^2_{\perp}/2)\nabla \ln B_0 +v^2_{\parallel}\mathbf{b}\cdot\nabla\mathbf{b}]$ is the magnetic drift velocity, $\langle A\rangle_{\alpha}$ denotes the gyro-phase averaging of $A$, $\langle \delta\mathbf{u}_{gj}\rangle_{\alpha}=(c/B_0)\mathbf{b}\times\nabla \langle \exp(\bm{\rho}_j\cdot\nabla)\delta L\rangle_{\alpha}$, $\delta L=\delta\phi-v_{\parallel}\delta A_{\parallel}/c$, and $\omega_{*j}=-i (cT/eB_0)_j\mathbf{b}\times \nabla \ln N_j\cdot\nabla$.  The governing field equations, meanwhile, are the quasi-neutrality condition
\begin{eqnarray}
\sum_{j=e,i}[(N_0e^2/T)_j\delta\phi - e_j\langle (J_k\delta g)_j\rangle_v]=0,\label{eq:QN}
\end{eqnarray}
and, for Alfv\'en waves, the nonlinear gyrokinetic vorticity equation in the wave-vector form \cite{LChenNF2001}
\begin{eqnarray}
&&i k_{\parallel} \delta J_{\parallel k} +(N_0e^2/T)_j(1-\Gamma_k)(\partial_t+i\omega_{*i})_k\delta \phi_k\nonumber\nonumber\\
&&-\sum_j\langle e_j J_k\omega_d\delta g_j \rangle=\sum_{\mathbf{k}=\mathbf{k}'+\mathbf{k}''} \Lambda^{k'}_{k''}\left\{\delta A_{\parallel k'}\delta J_{\parallel k''}/c\right.\nonumber\\
&&\hspace*{5em}\left.- e_i \langle (J_kJ_{k'}-J_{k''})\delta L_{k'}\delta g_{k''i}\rangle_v\right\}.\label{eq:vorticity}
\end{eqnarray}

We note that, in equations (\ref{eq:QN}) and (\ref{eq:vorticity}), $\mathbf{k}$ should be strictly understood as an operator; i.e., $\mathbf{k}=-i\nabla$, $ik_{\parallel}=\mathbf{b}\cdot\nabla$, and $k^2_{\perp}=-\nabla^2_{\perp}$, $\langle\cdots\rangle_v=\int d^3\mathbf{v}(\cdots)$, $J_k=J_0(k_{\perp}\rho)$, $\Gamma_k=I_0(b_k)\exp(-b_k)$, $b_k=k^2_{\perp}\rho^2_i$, $\rho_i=v_{ti}/\Omega_i$ with  $v_{ti}=\sqrt{T_i/m_i}$, $I_0$ being the modified Bessel function, and $J_0(k_{\perp}\rho_e)\simeq 1$ since $|k_{\perp}\rho_e|^2\ll1$.
In equation (\ref{eq:vorticity}), $\delta J_{\parallel}$ satisfies the Amp\'ere's law $\delta J_{\parallel}=-(c/4\pi) \nabla^2_{\perp}\delta A_{\parallel}$ and $\omega_d=\mathbf{v}_d\cdot\mathbf{k}_{\perp}$. Meanwhile, on the right hand side of equation (\ref{eq:vorticity}), $\Lambda^{k'}_{k''}=(c/B_0) \mathbf{b}\cdot(\mathbf{k}''\times\mathbf{k}')$, and the two nonlinear terms correspond, respectively, to the Maxwell and generalized gyrokinetic ion Reynolds stresses.  Note also that, since the ambient DWs are electrostatic, the Maxwell stress does not contribute to the scattering of TAE and, hence, can be ignored in the present analysis.

Adopting the ballooning-mode representation \cite{JConnorPRL1978}, we let the ambient stationary DW fluctuation be $\delta\phi_{DW}=\sum_{n_s}(\delta\phi_{s}+c.c.)/2$, and
\begin{eqnarray}
\delta\phi_{s}&=&(T_e/e)A_{n_s}\exp(-i\omega_{n_s}t+in_s\xi)\nonumber\\
&&\times\sum_{m_s} \exp(-im_s\theta)\Phi_s(n_sq-m_s),\label{eq:ballooning_repre}
\end{eqnarray}
where $\xi$ and $\theta$ are, respectively, the toroidal and poloidal angle coordinates,  $q(r)$ is the safety factor, and $\int^{\infty}_{-\infty}|\Phi_s|^2 d(n_sq-m_s)=1$ is  the normalization. We now let $\Omega_0=(\omega_0,n_0)$ be the test TAE with frequency $\omega_0$ and toroidal mode number $n_0$.  The scattering would occur for each $\delta\phi_{s}$, denoted as $\Omega_s=(\omega_{n_s},n_s)$, and generate upper and lower sidebands denoted, respectively, as $\Omega_+=(\omega_+=\omega_0+\omega_{n_s}, n_+=n_0+n_s)$ and  $\Omega_-=(\omega_-=\omega_{n_s}-\omega_0, n_-=n_s-n_0)$. As will be shown later in the detailed analysis, the nonlinearly generated $\Omega_{\pm}$ quasimodes correspond to mode-converted high-$n$  KAWs  \cite{AHasegawaPoF1976}; which are Landau damped by, mainly, electrons and, as a consequence, the scattering will contribute to the damping of the test TAE.  The net damping rate is then given by the sum of all $\Omega_s$ over $n_s$.

First, let us investigate the $\Omega_+$ channel. Analysis for the $\Omega_-$ is similar. Noting that $|k_{\parallel} v_{te}|\gg |\omega_k|\gg|k_{\parallel}v_{ti}|$ for all the modes considered here, we then find, from equation (\ref{eq:nl_gk}), $\delta g_k=\delta g^{(1)}_k+\delta g^{(2)}_k$, where
\begin{eqnarray}
\delta g^{(1)}_{+i}\simeq \left(\frac{e}{T_i}\right) F_{Mi}\left(1-\frac{\omega_{*i}}{\omega}\right)_+ J_+\delta\phi_+,\label{eq:response_u_l}
\end{eqnarray}
and
\begin{eqnarray}
\delta g^{(1)}_{+e}\simeq  - \left(\frac{e}{T_e}\right) F_{Me}\left(1-\frac{\omega_{*e}}{\omega}\right)_+ \delta\psi_+,
\end{eqnarray}
are the linear responses, and $\delta\psi_k=(\omega\delta A_{\parallel}/ck_{\parallel})_k$ is the effective potential due to the $\partial_t\delta A_{\parallel}/c$ induced parallel electric field. The nonlinear responses, $\delta g^{(2)}_k$, meanwhile, are given by
\begin{eqnarray}
\delta g^{(2)}_{+i}\simeq -i (\Lambda^s_0/2\omega_0) J_0J_s \frac{e}{T_i}F_{Mi}\left(\frac{\omega_{*i}}{\omega}\right)_s \delta\phi_s\delta\phi_0,\label{eq:response_u_nl}
\end{eqnarray}
with $\Lambda^s_0=(c/B_0)\mathbf{b}\cdot(\mathbf{k}_{0\perp}\times \mathbf{k}_{s\perp})$, and, noting $\Omega_s$ being an electrostatic mode, $\delta g^{(2)}_{+e}\simeq 0$.  Applying the quasi-neutrality condition, equation (\ref{eq:QN}), we then obtain
\begin{eqnarray}
\delta\psi_+=\sigma_{*+}\delta\phi_++i(\Lambda^s_0/2\omega_0)D_+\delta\phi_0\delta\phi_s,\label{eq:qn_u}
\end{eqnarray}
where
\begin{eqnarray}
\sigma_{*k}&=&[1+\tau-\tau \Gamma_k(1-\omega_{*i}/\omega)_k]/(1-\omega_{*e}/\omega)_k,\\
D_+&=&\tau(\omega_{*i}/\omega)_sF_+/(1-\omega_{*e}/\omega)_+,
\end{eqnarray}
$\tau=T_e/T_i$, and $F_+=\langle J_+J_0J_s F_{Mi}/N_0\rangle_v$.   Applying the nonlinear gyrokinetic vorticity equation, equation (\ref{eq:vorticity}), meanwhile, yields, after straightforward algebra,
\begin{eqnarray}
&&\tau b_+\left[\left(1-\frac{\omega_{*i}}{\omega}\right)_+\frac{(1-\Gamma_+)}{b_+}\delta\phi_+-\left(\frac{V^2_A}{b}\frac{k_{\parallel}bk_{\parallel}}{\omega^2}\right)_+\delta\psi_+\right]\nonumber\\
&=&-i(\Lambda^s_0/2\omega_0) \gamma_+\delta\phi_s\delta\phi_0,\label{eq:vorticity_u}
\end{eqnarray}
where $V_A=B_0/\sqrt{4\pi N_0m_i}$ is the Alfv\'en speed, and
\begin{eqnarray}
\gamma_+=\tau [\Gamma_s-\Gamma_0 + (\omega_{*i}/\omega)_s(F_+-\Gamma_s)].
\end{eqnarray}
Combining equations (\ref{eq:qn_u}) and (\ref{eq:vorticity_u}) then leads to the desired equation demonstrating the nonlinear generation of $\Omega_+$ quasimode by $\Omega_s$ and $\Omega_0$;
\begin{eqnarray}
\tau b_+\epsilon_{A+}\delta\phi_+=-i(\Lambda^s_0/2\omega_0) \beta_+ \delta\phi_s\delta\phi_0,\label{eq:dr_kaw_u}
\end{eqnarray}
where
\begin{eqnarray}
\epsilon_{Ak}=\left(1-\frac{\omega_{*i}}{\omega}\right)_k\frac{(1-\Gamma_k)}{b_k} -\left(\frac{V^2_A}{b}\frac{k_{\parallel}bk_{\parallel}}{\omega^2}\right)_k\sigma_{*k} \label{eq:saw_dr_wkb}
\end{eqnarray}
is the linear SAW/KAW operator, and
\begin{eqnarray}
\beta_+&=&\tau(\Gamma_s-\Gamma_0)
 +\tau \left(\frac{\omega_{*i}}{\omega}\right)_s \left[F_+-\Gamma_s-\right.\nonumber\\
 &+&\left.\left(\frac{V^2_A}{b}\frac{k_{\parallel}bk_{\parallel}}{\omega^2}\right)_+ \frac{\tau b_+ F_+}{(1-\omega_{*e}/\omega)_+} \right].
\end{eqnarray}

Next, we consider the feeding back to $\Omega_0$ via nonlinear coupling between $\Omega_+$ and $\Omega^*_s$.  Note that   the test TAE $\Omega_0$ evolution is also affected, on the same footing, by the other channel of $\Omega_s$ and $\Omega^*_-$ coupling, and the treatment is similar to  the  analysis of $\Omega_+$ and $\Omega^*_s$ couplings  presented in the following  equations (\ref{eq:response_0_nl}) to (\ref{eq:spatial_resonance}). The $\Omega_-$ contribution will be added in   equation (\ref{eq:TAE_nl_dr}).  Noting that $\delta g_{+i}=\delta g^{(1)}_{+i}+\delta g^{(2)}_{+i}$ given by equations (\ref{eq:response_u_l}) and (\ref{eq:response_u_nl}), we readily find
\begin{eqnarray}
\delta g^{(2)}_{0i}&\simeq& \left(\frac{e}{T}\right)_i F_{Mi}\left(\frac{\omega_{*i}}{\omega}\right)_s \left[i(\Lambda^s_0/2\omega_0) J_sJ_+\delta \phi^*_s\delta\phi_+\right.\nonumber\\
&&\left. +(\Lambda^s_0/2\omega_0)^2 J_0J^2_s|\delta\phi_s|^2\delta\phi_0 \right]. \label{eq:response_0_nl}
\end{eqnarray}
The second nonlinear term in equation (\ref{eq:response_0_nl}) is due to $\delta g^{(2)}_{+i}$, and may be regarded as the diagonal nonlinear term. The quasi-neutrality condition then yields
\begin{eqnarray}
\delta\psi_0=\left(\sigma_{*0}+\alpha_0|\delta\phi_s|^2\right)\delta\phi_0 - i(\Lambda^s_0/2\omega_0) D^+_0\delta\phi^*_s\delta\phi_+,\label{eq:qn_0_nl}
\end{eqnarray}
where
\begin{eqnarray}
D^+_0=\tau(\omega_{*i}/\omega)_s F_+/(1-\omega_{*e}/\omega)_0,
\end{eqnarray}
and $\alpha_0=-(\Lambda^s_0/2\omega_0)^2  \tau  (\omega_{*i}/\omega)_sF_2$ with $F_2\equiv \langle J^2_0J^2_s F_{Mi}/N_0\rangle_v$ leads to negligible nonlinear frequency shift. The nonlinear gyrokinetic vorticity equation, equation (\ref{eq:vorticity}), meanwhile, yields,
\begin{eqnarray}
&&\tau b_0 \left\{\left[\left(1-\frac{\omega_{*i}}{\omega}\right)_0\frac{(1-\Gamma_0)}{b_0} +\alpha^+_0|\delta\phi_s|^2 \right]\delta\phi_0 \right.\nonumber\\
&&\left.-\left(\frac{V^2_A}{b} \frac{k_{\parallel}bk_{\parallel}}{\omega^2}\right)_0\delta\psi_0\right\} = i\left(\frac{\Lambda^s_0}{2\omega_0}\right)\gamma^+_0 \delta\psi^*_s\delta\phi_+,\label{eq:vorticity_0_nl}
\end{eqnarray}
where
\begin{eqnarray}
\gamma^+_0 = \tau\left[\Gamma_s-\Gamma_+ + (\omega_{*i}/\omega)_s(F_+-F_s)\right],
\end{eqnarray}
 and $\alpha^+_0=-(\Lambda^s_0/2\omega_0)^2  (\omega_{*i}/\omega)_s(F_2-F_+)/b_0$  also leads to negligible nonlinear frequency shift. Combining equations (\ref{eq:qn_0_nl}) and (\ref{eq:vorticity_0_nl}) and neglecting the nonlinear frequency shift, we obtain the desired equation
\begin{eqnarray}
\tau b_0\epsilon_{A0}\delta\phi_0=i(\Lambda^s_0/2\omega_0)\beta^+_0\delta\phi^*_s\delta\phi_+, \label{eq:dr_tae_0}
\end{eqnarray}
where
\begin{eqnarray}
\beta^+_0\simeq (\sigma_{*0}\sigma_{*+}-\sigma_sF_+/\Gamma_s)/\sigma_{*0}\equiv \hat{\beta}_+/\sigma_{*0}.\label{eq:beta_u}
\end{eqnarray}
In deriving equation (\ref{eq:beta_u}), we have noted, in the nonlinear analysis, that $\Omega_s$ and $\Omega_0$ are normal modes; such that $\tau(\omega_{*i}/\omega)_s\simeq -(1+\tau-\tau\Gamma_s)/\Gamma_s$ and $|\epsilon_{A0}\delta\phi_0|\simeq O(|\delta\phi_s|^2)|\delta\phi_0|$.

Equations (\ref{eq:dr_kaw_u}) and (\ref{eq:dr_tae_0}) are the coupled equations for $\delta\phi_0$ and $\delta\phi_+$; which yield, formally,
\begin{eqnarray}
\tau b_0\epsilon_{A0}\delta\phi_0=\left[\left(\frac{\Lambda^s_0}{2\omega_0}\right)^2\beta^+_0\delta\phi^*_s\frac{\beta_+}{\left(\tau b_+\epsilon_{A+}\right)}\delta\phi_s\right]\delta\phi_0. \label{eq:nl_dr_wkb}
\end{eqnarray}

Equation (\ref{eq:nl_dr_wkb}) can be solved analytically by employing the scale separation between $\delta\phi_0$ and $\delta\phi_s$. More specifically, as TAE is typically excited by EPs, one has $k_{0\theta}\rho_{EP}=(n_0q/r)\rho_{EP}\sim O(1)$ and, for DWs, $k_{s\theta}\rho_i\sim O(1)$; and, hence, $|n_0/n_s|\sim O(\rho_i/\rho_{EP})\sim O(|T_i/T_{EP}|^{1/2})\ll1$. Denoting $\mathbf{x}_s=(R/n_s, r/m_s, 1/n_sq')$ and $\mathbf{x}_0=(R/n_0, r/m_0, 1/n_0 q')$ as, respectively, the microscopic DW and macro/mesoscopic TAE scales, and expanding $\delta\phi_0=\Phi_0(\mathbf{x}_0) + \tilde{\Phi}_0(\mathbf{x}_s,\mathbf{x}_0)$ with $|\tilde{\Phi}_0|/|\Phi_0|\sim O(|\delta\phi_s|^2)\ll1$, equation (\ref{eq:nl_dr_wkb}) then becomes, after averaging over the $\mathbf{x}_s$ scale,
\begin{eqnarray}
\tau b_0\epsilon_{A0}\Phi_0=\left\langle \left(\frac{\Lambda^s_0}{2\omega_0}\right)^2\beta^+_0\delta\phi^*_s\frac{\beta_+}{\tau b_+\epsilon_{A+}}\delta\phi_s\right\rangle_s\Phi_0, \label{eq:nl_dr_global}
\end{eqnarray}
where noting equation (\ref{eq:ballooning_repre}) and denoting $z_s=n_s q-m_s$ as the dimensionless   microscopic DW radial coordinate,
\begin{eqnarray}
&&\langle (\cdots)|\delta\phi_s|^2\rangle_s\nonumber\\
&=&\left(\frac{T_e}{e}\right)^2 |A_{n_s}|^2\int^{1/2}_{-1/2} d z_s  (\cdots)\sum_{m_s}|\Phi_s(z_s)|^2 \nonumber\\
&=&\left(\frac{T_e}{e}\right)^2 |A_{n_s}|^2 \int^{\infty}_{-\infty} d z_s (\cdots) |\Phi_s (z_s)|^2.
\end{eqnarray}

Focusing, furthermore, on the stability due to $\mbox{Im}(1/\epsilon_{A+})$ which can be formally expressed as   $\mbox{Im}(1/\epsilon_{A+})=-\pi\delta (\epsilon_{A+})$.  Noting also that, in equation (\ref{eq:nl_dr_global}), $\beta_+\delta(\epsilon_{A+})=(\hat{\beta}_+/\sigma_{*+})\delta(\epsilon_{A+})$ and $\hat{\beta}_+$ given by equation (\ref{eq:beta_u}). Applying the two-scale expansion; i.e.,  $k^2_{+\perp}\simeq k^2_{s\perp} + 2\mathbf{k}_{s\perp}\cdot \mathbf{k}_{0\perp} \simeq k^2_{s\perp} + 2 k_{s r} k_{0r}$,  to $\hat{\beta}_+$, we obtain
\begin{eqnarray}
&&\hat{\beta}_+\delta\phi_s\simeq \frac{T_e}{e} A_{n_s}e^{i(n_s\xi-\omega_{n_s} t)}\sum_{m_s}e^{-im_s\theta} \left(\tau+\frac{\sigma_s}{2\Gamma_s}\right)\nonumber\\
&& \times(\partial \Gamma_s/\partial b_s)(2 i k_{0r}\rho_i)(k_{s\theta}\rho_i)\hat{s}\partial\Phi_s/\partial z_s,\label{eq:beta_u_global}
\end{eqnarray}
and $\hat{s}=rq'/q$.  Meanwhile, from equation (\ref{eq:beta_u}),  $\beta^+_0\delta\phi^*_s=[\hat{\beta}_+\delta\phi_s]^*/\sigma_{*0}$.  We remark that,  $\mbox{Im}(1/\epsilon_{A+})=-\pi\delta (\epsilon_{A+})$ physically corresponds to absorption of the mode-converted $\Omega_+$ KAW via electron Landau damping     and, since the absorption occurs predominantly in a narrow region near the SAW resonance layer \cite{AHasegawaPoF1976},   its absorption rate can be quantitatively estimated by the Aflv\'en resonance absorption \cite{LChenPoF1974,FZoncaPoFB1993,FZoncaPoP1996};
\begin{eqnarray}
\mbox{Im}(1/\epsilon_{A+})=-\pi\delta (\epsilon_{A+})\simeq -(\pi/4 \sigma_{*+})\delta (z^2_s-z^2_+).\label{eq:spatial_resonance}
\end{eqnarray}
Here, we have noted $\epsilon_{A+}$ given by equation (\ref{eq:saw_dr_wkb}),  $k_{+\parallel}\simeq k_{s\parallel}=(n_sq-m_s)/qR=z_s/qR$ and $z^2_+=(1-\omega_{*i}/\omega)_+(1-\Gamma_+) (\omega/\omega_A)^2_+/(b_+\sigma_{*+})$, $\omega_A=V_A/qR$, $(\omega/\omega_A)_+\simeq 1/2$, and $b_+\simeq b_s\simeq b_{s\theta}$ for typical moderately to strongly ballooning DWs.  Substituting equations (\ref{eq:beta_u_global}) and (\ref{eq:spatial_resonance}) into equation (\ref{eq:nl_dr_global}),  and summing over all $\Omega_s$ over $n_s$, the TAE equation then becomes
\begin{eqnarray}
\tau b_0 \left[\epsilon_{A0}+i\nu(k_{0r}\rho_i)^2\right]\Phi_0=0,\label{eq:TAE_nl_dr}
\end{eqnarray}
where $\nu=\nu_++\nu_-$, and
\begin{eqnarray}
\nu_{\pm}&\simeq&  \pi  \left(\frac{\Omega_{ci}}{\omega_0}\right)^2 \sum_{n_s} |A_{n_s}|^2 \left[\left(\tau+\frac{\sigma_s}{2\Gamma_s}\right)\frac{\partial\Gamma_s}{\partial b_{s\theta}}\right]^2 \nonumber\\
&&\times b_{s\theta}\hat{s}^2\left(\sigma^2_{s\pm}z_{\pm}\right)^{-1}|\partial\Phi_s/\partial z_s|^2_{z_{\pm}}.\label{eq:nu}
\end{eqnarray}
In deriving equation (\ref{eq:TAE_nl_dr}), we have noted, again, that while the $\Omega_+$ channel results in the $\nu_+$ damping, the analysis of the $\Omega_-$ channel is similar and yields the $\nu_-$ term.

Equation (\ref{eq:TAE_nl_dr}) can be readily solved perturbatively in the ballooning space, $\eta$. Let $\hat{\Phi}(\eta)$ be the lowest-order eigenmode; i.e., $\hat{b}_0\hat{\epsilon}_{A0}(\eta,\partial_{\eta},\omega_{0})\hat{\Phi}_0(\eta)=0$ and $\omega_0=\omega_{0r}+i\gamma_{AD}$, with $\gamma_{AD}$ being the TAE damping rate induced by DW scattering. Here, from equation (\ref{eq:saw_dr_wkb}),
\begin{eqnarray}
\hat{b}_0\hat{\epsilon}_{A0}= \hat{b}_0-\left(\frac{\omega_A}{\omega_0}\right)^2\partial_{\eta} \hat{b}_0 (1+2\epsilon_0\cos\eta)\partial_{\eta},\nonumber
\end{eqnarray}
  $\hat{b}_0=(1+\hat{s}^2\eta^2)$,  $\epsilon_0=2(\Delta'+r/R)$ and $\Delta'\sim r/R$ is  the radial derivative of the Shafranov shift. Equation (\ref{eq:TAE_nl_dr}) then gives
\begin{eqnarray}
\frac{2\gamma_{AD}}{\omega_{0r}} \left\langle \hat{\Phi}_0\hat{b}_0\hat{\Phi}_0\right\rangle_{\eta}=-\left\langle \hat{\Phi}_0\hat{b}_0\nu b_{0\theta}\hat{s}^2\eta^2\hat{\Phi}_0\right\rangle_{\eta}. \label{eq:TAE_nl_ballooning}
\end{eqnarray}
Here, $\langle A\rangle_{\eta}=\int^{\infty}_{-\infty} A d\eta$. Given that, for TAE \cite{CZChengAP1985}, $\hat{\Phi}_{\eta}=[A\cos(\eta/2)+B\sin(\eta/2)]\exp(-\lambda |\eta|)/\hat{b}^{1/2}_0$,  and $\lambda^2=[(1+\epsilon_0)/4-(\omega_0/\omega_A)^2][(\omega_0/\omega_A)^2-(1-\epsilon_0)/4]\sim O(\epsilon^2)$; equation (\ref{eq:TAE_nl_ballooning}) then yields
\begin{eqnarray}
\frac{\gamma_{AD}}{\omega_{0r}}=-\frac{1}{4}\frac{\nu b_{0\theta}\hat{s}^2}{\lambda^2}.\label{eq:TAE_dr_global}
\end{eqnarray}

Equation (\ref{eq:TAE_dr_global}) is the desired analytic expression for TAE damping rate due to scattering by DWs. To obtain a quantitative estimate of $\gamma_{AD}$, we take $|\Phi_s(z_s)|=(\pi^{-1/4}\Delta^{-1/2}_s)\exp(-z^2_s/2\Delta^2_s)$ with $\Delta_s\gtrsim 1$ such that neighboring poloidal harmonics overlap to produce the ballooning mode structure. We then have, in equation (\ref{eq:nu}),  $|\partial\Phi_s/\partial z_s|^2\simeq (z^2_{\pm}/\pi^{1/2}\Delta^5_s)$;  noting $|z_{\pm}|<1/2$ and $\Delta_s\gtrsim 1$. Taking typical parameters, $|\Omega_{ci}/\omega_0|\sim O(10^2)$, $\sum_{n_s} |A_{n_s}|^2\simeq |e\delta\phi_{DW}/T_e|^2 \simeq |\delta n_{DW}/N_0|^2\sim O(10^{-4})$, $b_{s\theta}\sim\hat{s}\sim\tau\sim O(1)$, $4\lambda^2\sim O(\epsilon^2)\sim O(10^{-1} - 10^{-2})$, and $|z_{\pm}|/\Delta^5_s\sim O(10^{-1})$, we find  $(\gamma_{AD}/\omega_{0r})\simeq -O(1-10) b_{0\theta}$. Here, $\delta n_{DW}$ is the DW induced density fluctuation. As the EP drive maximizes around $k_{0\theta}\rho_{EP}\sim O(1)$, we have $b_{0\theta}\simeq (k_{0\theta}\rho_{EP})^2(T_i/T_{EP})\simeq T_i/T_{EP}$ and for, typically, $T_i/T_{EP}\sim O(10^{-2})$, $(\gamma_{AD}/\omega_{0r})\simeq -O(10^{-2}-10^{-1})$. We note that theoretical studies have shown that TAE instabilities excited by EPs have typically growth rates  $\gamma_{EP}/\omega_{0r}\sim O(10^{-2})$ \cite{LChenRMP2016}, $\gamma_{AD}$, thus, could be generally comparable to $\gamma_{EP}$, and, consequently, could significantly reduce or even, with sufficiently large DW intensity, suppress TAE fluctuations.  This raises the interesting implication that, in a burning plasma, with the presence of microscopic DW  turbulence, the critical gradient of $\alpha$-particles could be upshifted; leading to improved $\alpha$-particle confinement and, thereby, enhanced thermal plasma heating. Another important implication is that, as the high-$n$ KAW quasi-modes are dissipated, mainly, by electrons, the scattering and damping processes discussed here could provide a collisionless channel of transferring $\alpha$-particle energy to electrons via excitations of TAEs. Indeed, assuming the dissipated TAE wave energy is absorbed by electrons, we readily find that
\begin{eqnarray}
(d\beta_{e}/dt)_{AD}&=&4|\gamma_{AD}| |\delta B_{\perp}/B_0|^2\nonumber\\
&\simeq& O(10^{-2}-10^{-1})4\omega_{0r}|\delta B_{\perp}/B_0|^2.
\end{eqnarray}
Taking, typically, $|\delta B_{\perp}/B_0|\simeq 5\times 10^{-4}$ \cite{WHeidbrinkPRL2007} and $\omega_{0r}\simeq 10^6/s$, we have, $(d\beta_{e}/dt)_{AD}\simeq O(10^{-2}-10^{-1}s^{-1})$. As a comparison, electron heating by EPs ($\alpha$-particles)  with, e.g., a slowing down time $\tau_{SD}\sim O(10^2 \mbox{ms})$, is $(d\beta_e/dt)_{EP}\simeq O(10s^{-1})\beta_{EP}\simeq O(1-10s^{-1})\beta_e\simeq O(10^{-2}-10^{-1}s^{-1})$. Here, we have assumed $\beta_{EP}\sim O(\tau_{SD}/\tau_E)\beta_e$ with $\tau_E$ being the electron energy confinement time  is, typically, a fraction of $\beta_e$ and, again, $\beta_e\sim O(10^{-2})$.  These estimates suggest that $(d\beta_{e}/dt)_{AD}$ could, potentially, contribute significantly to ``anomalous" electron heating in burning plasmas.

In conclusion, we have employed the nonlinear gyrokinetic theory and investigated analytically the effects of ambient stationary DW turbulence on the linear stability of TAE via direct nonlinear wave-wave interactions. Our analysis demonstrates that the scattering of TAE by DWs could lead to appreciable damping of TAE due to electron Landau damping of the nonlinearly generated high-$n$ KAW quasimodes.  A corresponding analytical expression of the damping rate is derived and, for typical  parameters, the predicted damping rate could be comparable to the TAE growth rate driven by EPs. Our result, thus, suggests not only an additional TAE damping mechanism; but also carries interesting implications to improved $\alpha$-particle confinement as well enhanced collisionless heating of electrons in burning plasmas. Finally, we remark that, as noted earlier, our analysis adopts a paradigm electron DW model, where we keep nonuniform densities but assume uniform temperatures. It is obviously desirable to extend the present analysis to include finite temperature gradients and possible trapped-particle effects, as well as to other types of AEs; e.g., reversed shear Alfv\'en eigenmode (RSAE) \cite{SSharapovPLA2001,FZoncaPoP2002} and beta-induced Alfv\'en eigenmode (BAE) \cite{WHeidbrinkPRL1993,FZoncaPPCF1996}. These and other possible extensions are currently under investigation and will be reported in the future.

This work is supported by the National Key Research and Development  Program of China under Grant No. 2017YFE0301900, National Science Foundation of China under Grant Nos. 11235009 and 11875233, and ``Users of Excellence program of Hefei Science Center CAS under Contract No. 2021HSC-UE016".  This work has been carried out within the framework of the EUROfusion Consortium, funded by the European Union via the Euratom Research and Training Programme (Grant Agreement No 101052200 - EUROfusion). Views and opinions expressed are however those of the author(s) only and do not necessarily reflect those of the European Union or the European Commission. Neither the European Union nor the European Commission can be held responsible for them.

\end{document}